\documentclass[prd,twocolumn,showpacs,superscriptaddress,nofootinbib,floatfix,showkeys,10pt]{revtex4-2}
\usepackage{bm}
\usepackage{times}
\usepackage{braket}
\usepackage{amsfonts,amssymb,stmaryrd,latexsym,amsmath}
\usepackage[usenames,dvipsnames]{color}
\usepackage{epsfig}
\usepackage{slashed}
\usepackage{hyperref}
\usepackage{subfigure}
\usepackage{graphicx}
\usepackage{slashed}
\usepackage{orcidlink}
\usepackage{multirow}
\usepackage{dashrule}
\usepackage{bbm}
\usepackage[normalem]{ulem}
\usepackage{mdframed}

\usepackage{enumerate}
\usepackage{longtable}
\usepackage{booktabs}
\usepackage{makecell}

\allowdisplaybreaks[1]

\definecolor{nicered}{rgb}{0.7,0.1,0.1}
\definecolor{nicegreen}{rgb}{0.1,0.5,0.1}
\definecolor{emph}{rgb}{1,0,0}
\definecolor{doub}{rgb}{0.7,0.2,1.0}%{0.93,0.51,0.93}
\definecolor{navyblue}{RGB}{0, 110, 184}
\hypersetup{colorlinks,citecolor=nicegreen,linkcolor=nicered,urlcolor=navyblue}

\begin{document}

% %%%%%%%%%%%%%%%open the reply mode%%%%%%%%%%%%%%%%%%
% \onecolumngrid
% \input{reply.tex}
% \newpage
% \setcounter{page}{0}
% \setcounter{equation}{0} 
% \setcounter{table}{0} 
% \setcounter{figure}{0} 
% %%%%%%%%%%%%%%%open the reply mode%%%%%%%%%%%%%%%%%%
	
    \title{Doubly strange tetraquark as a $KK^*$ virtual state: quark-model study with harmonic traps}
	\author{Kai-Kai Zhang\,\orcidlink{0009-0003-3460-3483}}
    \affiliation{School of Physics, Southeast University, Nanjing 211189, China}
    \author{Zi-Yang Lin\,\orcidlink{0000-0001-7887-9391}}\email{linzy@seu.edu.cn}
    \affiliation{School of Physics, Southeast University, Nanjing 211189, China}
    \author{Yan-Ke Chen\,\orcidlink{0000-0002-9984-163X}}
	\affiliation{School of Physics and Center of High Energy Physics,
		Peking University, Beijing 100871, China}
	\author{Lu Meng\,\orcidlink{0000-0001-9791-7138}}\email{lmeng@seu.edu.cn}	
    \affiliation{School of Physics, Southeast University, Nanjing 211189, China}

	\begin{abstract}
We extract hadron--hadron scattering information from quark-level interactions using an artificial harmonic trap and a four-quark basis that includes both meson--meson and compact diquark--antidiquark configurations.
Combining the Busch--Englert--Rza\.{z}ewski--Wilkens formula with the effective range expansion (ERE), we determine phase shifts, scattering lengths, effective ranges, and pole positions from the trapped spectrum.
We control uncertainties from basis truncation, finite trap frequencies, truncation of ERE, and  quark-model parameters.
The extracted $DD^*$ and $BB^*$ binding energies agree with direct calculations, providing benchmarks.
For the doubly strange $T_{ss}$ with $I(J^P)=0(1^+)$, we predict a $KK^*$ virtual-state pole about 30--35~MeV below threshold in the stable-$K^*$ approximation.
Both the AL1 model and its modified version that better reproduces the measured $T_{cc}^+$ binding energy yield this result.
This finding differs from earlier predictions of a bound state and motivates searches for the $T_{ss}$ at facilities such as the High Intensity heavy Ion Accelerator Facility (HIAF). This framework could be extended to other hadron-hadron channels to extract scattering observables and pole positions.
\end{abstract}
	\maketitle

    \section{Introduction}

In 2003, the observation of the exotic hadron candidate $X(3872)$ by the Belle Collaboration \cite{Belle:2003nnu} sparked extensive experimental and theoretical investigations into exotic hadrons. In 2021, the LHCb Collaboration observed the doubly charmed tetraquark $T_{cc}^{+}$, subsequently characterizing it as a near-threshold $DD^*$ state \cite{LHCb:2021vvq,LHCb:2021auc}. Meanwhile, although not yet observed experimentally, a deeply bound $T_{bb}$ tetraquark state has been consistently predicted by various theoretical approaches; see, e.g., Refs.~\cite{Francis:2016hui,Eichten:2017ffp,Karliner:2017qjm,Lu:2020rog,Meng:2023jqk}. For comprehensive reviews of exotic hadrons, see Refs.~\cite{Jaffe:2004ph,Liu:2019zoy,Chen:2022asf,Liu:2024uxn,Guo:2017jvc,Meng:2022ozq,Wang:2025sic,Dai:2026fkg,Bai:2026atm,Brambilla:2019esw}. Given the established $T_{cc}^+$ state and the widely anticipated $T_{bb}$ state, the doubly strange $T_{ss}$ state (quark content $\bar{q}\bar{q}ss$ with $q=u,d$) warrants further exploration. Production mechanisms for doubly strange tetraquarks have been discussed in Ref.~\cite{Liu:2008ck}. Searches for the $T_{ss}$ could be pursued at facilities such as the High Intensity heavy Ion Accelerator Facility (HIAF)~\cite{An:2025lws}.

Investigations into doubly strange tetraquarks date back several decades \cite{Jaffe:1976ig,Jaffe:1976ih,Vijande:2003ki}. Following the 2003 report of the strange pentaquark $\Theta^+(1540)$ ($uudd\bar{s}$)~\cite{LEPS:2003wug}, several studies explored possible doubly strange tetraquark partners motivated by this observation~\cite{Karliner:2004sy,Burns:2004wy,Kanada-Enyo:2005gga,Cui:2005az}. Subsequent work investigated their spectra~\cite{Chen:2006hy,Wang:2007kb,Gao:2012zza} and production mechanisms~\cite{Liu:2008ck}. However, subsequent experimental searches did not confirm the $\Theta^+$ signal~\cite{BES:2004kia,Battaglieri:2005er,McKinnon:2006zv}. Recently, inspired by the observation of the $T_{cc}^+$, renewed interest has been sparked, leading to several new predictions for its strange partner, the $T_{ss}$~\cite{Wang:2024kke,Ji:2024znj,Liu:2025fpx,Tang:2025kkstar}. Among the various investigations of $T_{ss}$ states, the configuration with quantum numbers $I(J^{P})=0(1^+)$ has received particular attention~\cite{Jaffe:1976ig, Vijande:2003ki, Kanada-Enyo:2005gga, Cui:2005az, Vijande:2006jf, Wang:2007kb, Yang:2009zzp, Garcia-Recio:2010enl, Gao:2012zza, Wang:2024kke, Ji:2024znj, Liu:2025fpx,Tang:2025kkstar}. With these quantum numbers, the $T_{ss}$ is a natural strange partner of the $T_{cc}^+(3875)$ and couples to $KK^*$ in S wave.

Extensive quark-model studies of doubly heavy tetraquarks~\cite{Zouzou:1986qh,Semay:1994ht,Meng:2023jqk,Yang:2009zzp,Meng:2021yjr,Ma:2023int,Wu:2024zbx} motivate the study on $T_{ss}$. Existing calculations employ chiral quark models~\cite{Vijande:2003ki,Wang:2007kb,Yang:2009zzp,Gao:2012zza,Ji:2024znj}, including implementations based on Refs.~\cite{Vijande:2004he,Gonzalez:2012gka}, as well as the quark delocalization color screening model (QDCSM)~\cite{Yang:2009zzp,Liu:2025fpx} and Cornell-type diquark-antidiquark models~\cite{Lodha:2024yfn}. However, representative chiral-model calculations incorporating both diquark-antidiquark and dimeson configurations predict $T_{cc}$ binding energies of order $10^2$~MeV~\cite{Yang:2019itm,Meng:2023jqk}. This is in stark contrast to the experimental observation, which identifies the $T_{cc}^+$ as a near-threshold state. On the other hand, investigations using a minimal Cornell-like potential~\cite{Eichten:1978tg} predict a loosely bound state. For example, with the AL1 potential~\cite{Semay:1994ht,Silvestre-Brac:1996myf}, the $T_{cc}$ binding energy is approximately 15~MeV~\cite{Meng:2023jqk}. This motivates a four-body study of $KK^*$ scattering with a minimal Cornell-like quark interaction, without meson-exchange terms.

As indicated by studies on the quark-mass dependence of $QQ\bar{q}\bar{q}$ multiquark states~\cite{Zouzou:1986qh,Eichten:2017ffp,Collins:2024sfi}, given the $T_{cc}^+$ is a very near-threshold state, the $T_{ss}$ is more likely to be unbound even if the $KK^*$ interaction remains attractive (since $m_s < m_c$). As the attraction in an S-wave channel is reduced, a shallow bound-state pole can move onto the unphysical sheet and become a virtual state~\cite{Lin:2024qcq}. If such a virtual-state pole lies close to the threshold, it can still produce observable effects~\cite{Guo:2019twa}. Identifying such poles directly from four-quark dynamics remains computationally challenging. In recent years, computational techniques for multiquark states have advanced rapidly. In particular, the complex-scaling method allows resonance states to be identified directly in four-quark calculations~\cite{Chen:2023syh,Wu:2024euj,Wu:2024hrv}. Meanwhile, deep neural networks have also been applied to multiquark systems~\cite{Wu:2025wvv}. Virtual-state poles have been extracted in two-body calculations~\cite{Chen:2023eri} and in a coupled-channel quark-model calculation using the resonating group method~\cite{Ortega:2018cnm}.
The resonating group method uses fixed internal meson wave functions. However,
four-quark benchmark calculations show that omitting diquark--antidiquark configurations can lead to a substantial underestimate of tetraquark binding energies~\cite{Meng:2023jqk}.
Here we include both meson--meson and diquark--antidiquark configurations and extract meson--meson scattering information and a virtual-state pole from the discrete spectrum of a trapped four-quark Hamiltonian.

The artificial harmonic-oscillator (HO) trap method was originally proposed to investigate the scattering of cold atoms~\cite{Busch:1998cey}. In this framework, the so-called Busch--Englert--Rza\.{z}ewski--Wilkens (BERW) formula was established, which relates the scattering parameters to the discrete energy levels in the HO trap. Its prediction for the two-body binding energy was subsequently tested experimentally with ultracold atoms in optical-lattice wells~\cite{stoferleMoleculesFermionicAtoms2006}. Since then, the method has been systematically extended to higher partial-wave scattering~\cite{suzukiTwoatomEnergySpectrum2009}, charged two-body systems~\cite{Guo:2021qfu}, and coupled-channel systems~\cite{Guo:2021uig}. The standard BERW formula is exact for zero-range interactions; for finite-range interactions, it becomes exact in the weak-trap limit \(a_{\rm HO}\to\infty\), or equivalently, \(\omega\to0\). Here, \(a_{\rm HO}=(\mu\omega)^{-1/2}\) is the harmonic-oscillator length for the relative motion, with \(\omega\) and \(\mu\) denoting the trap frequency and reduced mass, respectively. To account for corrections arising from a finite trap frequency $\omega$, modified formulas have been developed~\cite{Luu:2010hw,Zhang:2019cai}. In addition to cold-atom physics, this method has also been widely applied to investigate nuclear scattering and reactions~\cite{Schafer:2022hzo,Yang:2025mhg,Yang:2026vrl,Zhang:2020rhz,Zhang:2024vmz}. Recent developments include coupled-channel scattering from artificial confinement~\cite{Attia:2026xtt}, data-driven trap theory~\cite{Zhang:2025ypt}, and applications to quantum computing~\cite{Yusf:2024igb}.

    \begin{figure}[t]
    \centering
    \includegraphics[width=0.7\linewidth]{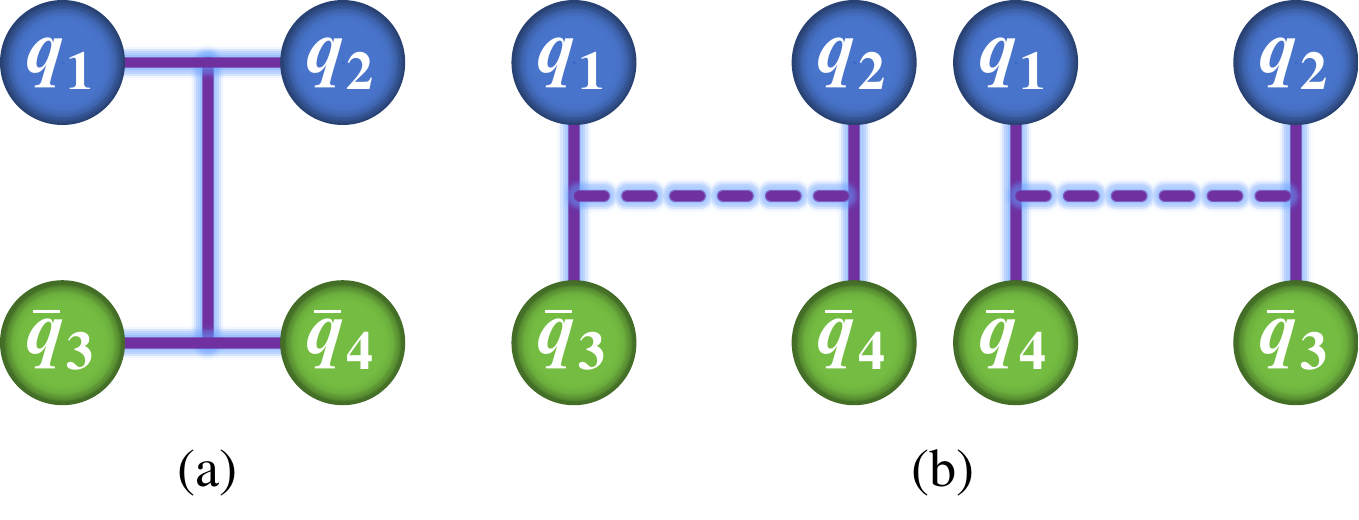}
    \caption{Jacobi sets: (a) diquark-antidiquark (${\rm jac}=1$) and (b) two meson--meson configurations (${\rm jac}=2,3$). The dotted lines denote the {\it scattering coordinates} describing the relative motion between the two mesons, while the solid lines denote the  {\it internal coordinates.}}
    \label{fig:jcb}
    \end{figure}

In this work, we employ the artificial HO trap method to investigate $KK^*$ scattering within a four-quark calculation, focusing on the $I(J^P)=0(1^+)$ channel. For a two-body system in a sufficiently weak harmonic trap, the Busch--Englert--Rza\.{z}ewski--Wilkens (BERW) formula reads~\cite{suzukiTwoatomEnergySpectrum2009}:
\begin{equation}
\begin{aligned}
p^{2l+1}\cot\delta_l(p)
= (-1)^{l+1}(4\mu\omega)^{l+1/2}
\frac{\Gamma\left(\frac34+\frac l2-\frac{\mathcal{E}(\omega)}{2\omega}\right)}
     {\Gamma\left(\frac14-\frac l2-\frac{\mathcal{E}(\omega)}{2\omega}\right)},
\end{aligned}
\label{eq:berw}
\end{equation}
where $l$ is the orbital angular momentum, $\delta_l(p)$ is the scattering phase shift, and $\mu$ is the reduced mass of the two-body system. The relative energy in the trap is given by $\mathcal{E}(\omega)=E(\omega)-E_{\text{thr}}(\omega)$, where $E(\omega)$ is the total energy obtained by diagonalizing the Hamiltonian, and $E_{\text{thr}}(\omega)$ is the corresponding threshold energy in the trap. The squared relative momentum and the oscillator length are defined as $p^2=2\mu\mathcal{E}(\omega)$ and $a_{\rm HO}=(\mu\omega)^{-1/2}$, respectively.

In practical calculations, we perform a four-quark calculation in the HO trap. The single-particle HO potentials can be decomposed into the center-of-mass and relative Jacobi coordinate HO traps:
\begin{equation}
\begin{aligned}
V_{\rm HO}
=
\frac{1}{2}\omega^2\sum_i m_i\mathbf{r}_i^2
=
\frac{1}{2} M\omega^2\mathbf{R}_{\rm CM}^2
+
\frac{1}{2}\omega^2\sum_{a=1}^{3}\mu_a\mathbf{x}_a^2,
\end{aligned}
\label{eq:vho}
\end{equation}
where $M$ is the total mass, $\mathbf{R}_{\rm CM}$ is the center-of-mass coordinate, and $\mathbf{x}_a$ are the Jacobi coordinates shown in Fig.~\ref{fig:jcb} with their corresponding reduced masses $\mu_a$. 

We choose the trap frequency $\omega$ such that the oscillator length associated with the internal coordinates, denoted as $b_{\rm HO}$, significantly exceeds the characteristic sizes of the $K^{(*)}$ mesons and of any compact tetraquark configuration, so that the trap perturbs their internal structures only weakly. At the same time, $\omega$ must be chosen such that the oscillator length $a_{\rm HO}$, associated with the scattering (meson--meson relative) coordinate, is much larger than the range of the intercluster interaction. Consequently, the trap potential effectively acts on the two meson clusters as a whole, reducing the four-quark system to an effective two-body problem of two interacting clusters confined within the trap. Thus, we can apply the two-body BERW formula to extract the $KK^*$ scattering phase shift directly from the four-quark energy spectrum. We benchmark this approach against direct bound-state calculations for the $DD^*$ and $BB^*$ systems.

    \section{Hamiltonian and wave functions}

    We employ the nonrelativistic AL1 two-body interaction potential~\cite{Semay:1994ht,Silvestre-Brac:1996myf}. The Hamiltonian in the harmonic trap is given by
\begin{equation}
\begin{aligned}
H = \sum_i \left( m_i + \frac{\mathbf{p}_i^2}{2m_i} \right) + V_{\rm HO} - T_{\rm CM} + \sum_{i<j} V_{ij},
\end{aligned}
\label{eq:hint-trap}
\end{equation}
where $m_i$ and $\mathbf{p}_i$ are the constituent mass and momentum of the $i$-th quark (or antiquark), respectively, and $V_{\rm HO}$ is the trap potential defined in Eq.~\eqref{eq:vho}. With the total mass $M=\sum_i m_i$ and the center-of-mass coordinate $\mathbf{R}_{\rm CM}=\sum_i m_i\mathbf{r}_i/M$, the term $T_{\rm CM} = (\sum_i \mathbf{p}_i)^2/(2M) + \frac{1}{2}M\omega^2\mathbf{R}_{\rm CM}^2$ represents the center-of-mass energy, which is subtracted to isolate the internal dynamics. For the two-body interaction, let $r_{ij}=|\mathbf{r}_i-\mathbf{r}_j|$ and $\mathcal{C}_{ij}=-\frac{3}{16}(\boldsymbol{\lambda}_i\cdot\boldsymbol{\lambda}_j)$. The AL1 potential consists of one-gluon-exchange (OGE) and confinement terms:
\begin{equation}
\begin{aligned}
V_{ij} &= \mathcal{C}_{ij}\left(V_{ij}^{\rm OGE}+V_{ij}^{\rm conf}\right), \quad V_{ij}^{\rm conf} = \lambda r_{ij} - \Lambda, \\
V_{ij}^{\rm OGE} &= -\frac{\kappa}{r_{ij}} + \frac{2\pi\kappa'}{3m_i m_j} \frac{e^{-r_{ij}^2/r_0^2}}{\pi^{3/2}r_0^3} (\boldsymbol{\sigma}_i\cdot\boldsymbol{\sigma}_j).
\end{aligned}
\label{eq:al1-potential}
\end{equation}
For antiquarks, the color matrices are taken as $-\boldsymbol{\lambda}_i^*$, and $\boldsymbol{\sigma}_i$ are the spin Pauli matrices. The pair-dependent smearing parameter is given by $r_0=A[2m_i m_j/(m_i+m_j)]^{-B}$. We denote the Coulomb term alone by $V^C_{ij}=-\mathcal{C}_{ij}\kappa/r_{ij}$. The AL1 parameters are taken from Ref.~\cite{Silvestre-Brac:1996myf}.

The trapped four-body problem is solved using the Gaussian expansion method (GEM); see Refs.~\cite{Hiyama:2003cu,Meng:2023jqk,Wu:2024zbx} for technical details. Here, we briefly outline the key steps of the calculation. For a tetraquark composed of two identical quarks ($q_1, q_2$) and two identical antiquarks ($\bar{q}_3, \bar{q}_4$) with quantum numbers $I(J^P)=0(1^+)$, the total wave function is constructed as
\begin{equation}
\Psi = \mathcal{A} \sum_\alpha C_\alpha \left( \Phi_\alpha \otimes \chi_f \otimes \chi_{s,a} \otimes \chi_{c,b} \right),
\label{eq:total-wf}
\end{equation}
where the index $\alpha=(\text{jac}, \mathbf{n}, a, b)$ specifies the Jacobi coordinate set $\text{jac}$, the spatial Gaussian indices $\mathbf{n}$, the spin state $a$, and the color state $b$. The antisymmetrization operator $\mathcal{A}=\frac{1}{2}(1-P_{12})(1-P_{34})$ ensures the correct Fermi-Dirac statistics for the identical quarks and antiquarks. The three Jacobi coordinate sets shown in Fig.~\ref{fig:jcb} comprise one diquark-antidiquark and two meson--meson configurations. For the $KK^*$ configuration in the $T_{ss}$ channel, the flavor wave function is given by
\begin{equation}
\chi_f = [nn]\bar{s}\bar{s}, \quad \text{with} \quad [nn] = \frac{ud-du}{\sqrt{2}}.
\end{equation}
The three spin-1 basis states are
\begin{equation}
\begin{aligned}
\chi_{s,1} &= [(q_1q_2)_1(\bar{q}_3\bar{q}_4)_0]_1, \\
\chi_{s,2} &= [(q_1q_2)_0(\bar{q}_3\bar{q}_4)_1]_1, \\
\chi_{s,3} &= [(q_1q_2)_1(\bar{q}_3\bar{q}_4)_1]_1,
\end{aligned}
\label{eq:spin-basis}
\end{equation}
and the two color-singlet basis states are
\begin{equation}
\begin{aligned}
\chi_{c,1} &= [(q_1q_2)^{\bar{\mathbf{3}}_c} (\bar{q}_3\bar{q}_4)^{\mathbf{3}_c}]_1, \\
\chi_{c,2} &= [(q_1q_2)^{\mathbf{6}_c} (\bar{q}_3\bar{q}_4)^{\bar{\mathbf{6}}_c}]_1.
\end{aligned}
\label{eq:color-basis}
\end{equation}
The spatial wave functions are constructed as products of Gaussian functions, one for each Jacobi coordinate:
\begin{equation}
    \Phi_\alpha = \prod_{k=1}^{3} \phi_{n_k},\quad \phi_n(\mathbf{r}) = \left(\frac{2}{\pi r_n^2}\right)^{3/4} e^{-r^2/r_n^2},
\end{equation}
where the range parameters follow a geometric progression.  In our calculations, we incorporate both diquark-antidiquark and dimeson spatial configurations to describe both compact tetraquark and $KK^*$ molecular states. We employ eight Gaussian basis functions for each internal coordinate. For the meson--meson relative coordinate, we use 22 Gaussian basis functions with range parameters (the largest one  $r_{\rm max}=5a_{\rm HO}$ ) on the two-segment grid specified below.

\section{Numerical Setup and Systematic Uncertainties}

We first investigate the existence of the doubly strange tetraquark with $I(J^{P})=0(1^+)$ using the GEM~\cite{Meng:2023jqk} without the HO trap potential. This calculation yields no bound state below the $KK^*$ threshold. Consequently, we resort to extracting the $KK^*$ scattering phase shift via the BERW formula in the presence of an artificial HO trap, employing the effective range expansion (ERE) to locate the possible virtual-state pole. To assess the reliability of this combined HO trap and ERE approach, we use the doubly charmed ($T_{cc}$) and doubly bottomed ($T_{bb}$) tetraquarks as benchmarks, as both systems support bound states in the AL1 model. Specifically, we compare the bound-state pole obtained directly from the standard basis expansion (without the trap) with the pole extracted using the HO trap method combined with the ERE. 

Furthermore, we have carefully optimized our numerical setup to strictly control the uncertainties associated with the three main steps of our analysis: 
\begin{itemize}
    \item the precision of the trapped energy levels, ensuring a reliable discrete spectrum; 
    \item the validity and accuracy of the BERW formula, guaranteeing that reliable scattering phase shifts can be extracted from these precise energy levels; 
    \item the convergence and accuracy of the ERE, ensuring that the pole position can be reliably determined from the extracted phase shifts.
\end{itemize}
 Below, we detail the specific numerical configurations and tests employed to constrain these systematic uncertainties.

      \begin{figure*}[t]
    \centering
    \includegraphics[width=\textwidth]{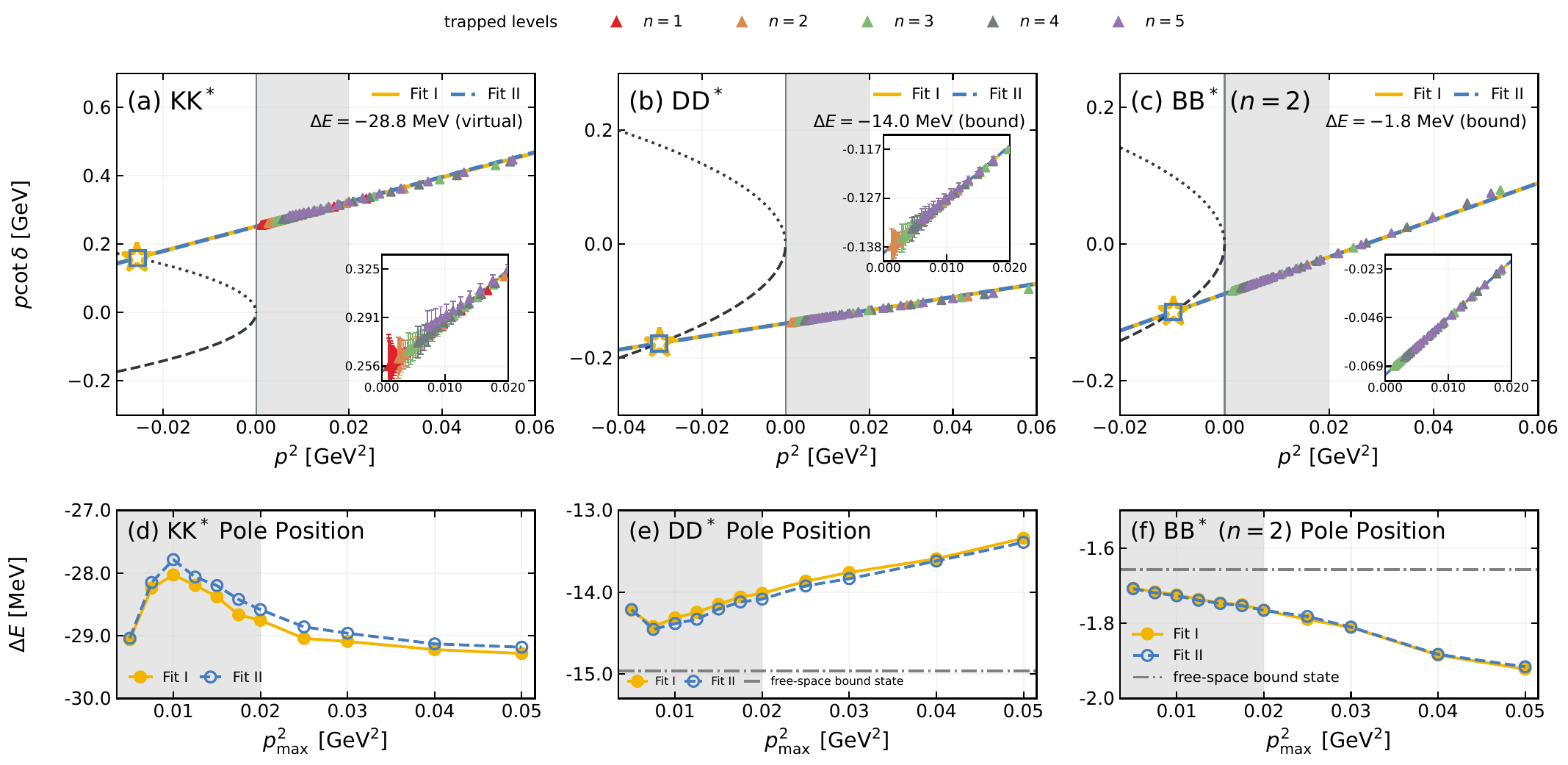}
    \caption{Phase shifts and pole positions extracted from the BERW formula for the $KK^*$, $DD^*$, and $BB^*$ systems in the AL1 quark model. The upper panels show the phase shifts and pole positions obtained from Fit~I (the standard ERE in Eq.~\eqref{eq:ere}) and Fit~II (the ERE including the leading finite-$\omega$ correction in Eq.~\eqref{eq:ereomega}). The error bars in $p \cot \delta$ are propagated from the basis truncation errors in the energy levels.  The lower panels display the dependence of the extracted pole positions on the choice of the $p^2$ fitting range, with the shaded region indicating our nominal choice. For the $DD^*$ system and the first excited state of $BB^*$, the binding energies obtained from the basis expansion are also shown for comparison. See the main text for more details. }
    \label{fig:pole-window-ere}
    \end{figure*}

In the calculation, the trap frequency $\omega$ must be chosen within an intermediate range to balance physical validity and numerical feasibility. On the physical side, $\omega$ should be selected such that the oscillator lengths associated with the internal coordinates, $b_{\rm HO}$, are large enough such that the internal structures of the quark--antiquark mesons are only slightly distorted, and the oscillator length associated with the scattering coordinate, $a_{\rm HO}$, is sufficiently large for the BERW formula to remain valid. On the numerical side, an excessively small $\omega$ would produce an overly dense discrete energy spectrum; exploring the energy range of interest would then require including a large number of highly excited states, significantly increasing the computational cost of the few-body calculation. Striking a balance between these considerations, we vary $\omega$ such that $a_{\rm HO}$ ranges from 2.0 to 10.0~fm. Within this range, the internal energy shifts due to HO traps of all six mesons---$K$, $K^{*}$, $D$, $D^{*}$, $B$, and $B^{*}$---remain below 1~MeV (see the Supplemental Material~\cite{Suppl}). To further reduce the uncertainties arising from the distortion of the quark--antiquark mesons, we use the meson--meson threshold evaluated at the same trap frequency $\omega$, rather than the free-space threshold, which effectively cancels the trap-induced energy shift of the two mesons.

In our calculations of the energy levels, we employ distinct Gaussian basis sets for the internal and scattering coordinates. For the internal coordinates, which describe the compact structures of the constituent mesons, we use 8 Gaussian basis functions for each coordinate, with range parameters spanning from 0.1 to 2.0~fm, achieving accurate results for the internal dynamics. To further reduce systematic uncertainties, we use the same 8 Gaussian basis functions to calculate the quark-antiquark meson masses, which are then substituted into the threshold energy $E_{\rm thr}(\omega)$ in Eq.~(\ref{eq:berw}). We have verified the accuracy of this basis set by comparing our results with those obtained using 100 basis functions, confirming that the finite-basis error is of the order of 0.1~keV.

In contrast, the scattering coordinate encodes the crucial scattering information and requires a more extended basis. To optimize the efficiency of this basis, we divide it into two segments with different geometric scaling parameters. Specifically, the first segment consists of 9 basis functions with range parameters ranging from 0.1 to $\max(2.0\text{ fm}, a_{\rm HO}/2)$. The second segment comprises 14 basis functions, with range parameters extending from $\max(2.0\text{ fm}, a_{\rm HO}/2)$ up to $5a_{\rm HO}$. Accounting for one shared basis function at the boundary between the two segments, we employ a total of $9 + 14 - 1 = 22$ Gaussian basis functions for this coordinate. This two-segment configuration ensures high accuracy in the short-range region while extending the basis to sufficiently large distances to properly capture the long-range scattering dynamics. To quantify the numerical uncertainty arising from the finite basis for the scattering coordinate, we estimate the energy level precision using a pure-harmonic-oscillator two-meson model. By comparing the numerical results obtained with the same 22 Gaussian basis functions against the analytical results, we determine an uncertainty of $5$~keV for both the $KK^*$ and $DD^*$ systems, and $1$~keV for the $BB^*$ system. Details of this estimation are given in the Supplemental Material~\cite{Suppl}. This uncertainty is then propagated through the subsequent steps of our analysis.

For each HO trap, we retain the first five energy levels ($n=1$--$5$). For each positive relative energy $\mathcal{E}_n(\omega)$, Eq.~\eqref{eq:berw} yields a data point $(p^2, p\cot\delta)$, which is presented in the upper panels of Fig.~\ref{fig:pole-window-ere}. The uncertainties of these points are propagated from the previously evaluated uncertainties of the energy levels. As observed for all three systems, these points lie approximately on a straight line. This indicates that the scattering amplitude can be well described by the ERE truncated at $\mathcal{O}(p^2)$: 
\begin{equation}
    p\cot\delta = -1/a_0 + \frac{1}{2}r_e p^2,~\label{eq:ere}
\end{equation}
which corresponds to Fit~I in Fig.~\ref{fig:pole-window-ere}. Although higher-order terms in the ERE could in principle be considered for the current data, it is evident from the figure that such higher-order effects are negligible. The intersection of the effective range function and the line $ip$ (i.e., solving $p\cot\delta(p) - ip = 0$) corresponds to the pole position. Specifically, a pole located at $p=i\kappa$ ($p=-i\kappa$) with $\kappa>0$ and $E_{\mathrm{pole}}=-\kappa^2/(2\mu)$ represents a bound (virtual) state on the physical (second) Riemann sheet.

A potential ambiguity in applying Eq.~\eqref{eq:berw}  to the {\it nonrelativistic} four-quark system as a two-cluster scattering problem concerns the choice of cluster masses: the sums of the constituent quark and antiquark masses or the meson bound-state masses. For a fixed trapped spectrum, this choice affects the relative momentum but cancels from the BERW relation when expressed in terms of the phase shift as a function of the relative energy, provided that the reduced mass is used consistently. The extracted pole energy is therefore independent of this choice. Accordingly, we report the pole position as an energy relative to the $KK^*$, $DD^*$ or $BB^*$ thresholds. The scattering length $a_0$ and effective range $r_e$ both scale as $\mu^{-1/2}$, so switching between the two mass conventions changes their values only modestly in the channels considered here.

Furthermore, we investigate the finite-$\omega$ effects by treating the data points in Fig.~\ref{fig:pole-window-ere} as results obtained at finite $\omega$~\cite{Luu:2010hw,Zhang:2019cai}. We extend the ERE to include a leading $\omega$-dependent term: 
\begin{equation}
    p\cot\delta(\omega) = -1/a_0 + \frac{1}{2}r_e p^2 + d_0 \omega^2,~\label{eq:ereomega}
\end{equation}
which is denoted as Fit~II in the figure. The physical observables are then extracted by taking the limit $\omega \to 0$. The pole energies extracted from Fit~I and Fit~II differ by less than $0.5~\mathrm{MeV}$, so the effect of the leading finite-$\omega$ correction can be neglected at the present precision.

In the lower panels of Fig.~\ref{fig:pole-window-ere}, we also examine the sensitivity of the results to the choice of the fitting range for $p^2$. The impact is found to be minimal. Ultimately, we select the data with $p^2 < 0.02~\mathrm{GeV}^2$ as our nominal fitting range. To estimate the overall systematic uncertainty of this approach, we benchmark the HO trap method combined with BERW+ERE against the direct basis expansion method for the $D D^*$ and $B B^*$ systems, both of which possess bound states. For the first excited state of the $B B^*$ system, which has a binding energy of approximately $1.65~\mathrm{MeV}$, the deviation of the HO trap method is less than $0.2~\mathrm{MeV}$. For the $D D^*$ system, with a binding energy of about $15~\mathrm{MeV}$, the deviation is less than $1~\mathrm{MeV}$. Based on these benchmarks, we can reasonably conclude that for the $K K^*$ system, the systematic deviation of this method should be at most a few $\mathrm{MeV}$.

    \section{Results and discussion}

With the numerical uncertainties quantified, we present the scattering lengths, effective ranges, and the nature of the near-threshold states (bound or virtual) for the $KK^*$, $DD^*$, and $BB^*$ systems, extracted via the BERW formula, in Table~\ref{tab:scattering-results}. The AL1 model predicts an attractive interaction in the $KK^*$ channel; however, it is not strong enough to form a bound state. Instead, a virtual state is predicted to lie approximately 30~MeV below the threshold. A comparison of the $DD^*$ and $BB^*$ scattering parameters with experimental and lattice results is provided in the Supplemental Material~\cite{Suppl}.

  \begin{table}[t]
    \caption{Low-energy properties of the $I(J^P)=0(1^+)$ channels in the AL1 quark model. The pole positions are given with respect to the corresponding thresholds. For the $DD^*$ system and the first excited state of $BB^*$, the binding energies obtained from the basis expansion, labeled as ``free space,'' are also shown for comparison. A dash (``$-$'') indicates the absence of a bound state in the $KK^*$ system. The uncertainties are propagated from the finite-basis truncation errors in the energy levels, which have been verified to be the dominant source of numerical uncertainty (see the main text for details).}
    \label{tab:scattering-results}
    \centering
    \setlength{\tabcolsep}{3pt}
    \begin{tabular*}{\columnwidth}{@{\extracolsep{\fill}}lccc@{}}
    \hline\hline
    Channel & $KK^*$ & $DD^*$ & $BB^* (n=2)$ \\
    \hline
    $a_0$ (fm)
    & $-0.783^{+0.030}_{-0.030}$
    & $1.421^{+0.071}_{-0.070}$
    & $2.721^{+0.043}_{-0.043}$ \\[3pt]
    $r_e$ (fm)
    & $1.415^{+0.190}_{-0.211}$
    & $0.455^{+0.165}_{-0.146}$
    & $1.065^{+0.023}_{-0.023}$ \\[3pt]
    BERW (MeV)
    & $-28.75^{+3.35}_{-4.42}$
    & $-14.02^{+3.57}_{-9.97}$
    & $-1.766^{+0.122}_{-0.143}$ \\[3pt]
    Pole type & virtual & bound & bound \\
    Free space (MeV) & - & $-15.0$ & $-1.7$ \\
    \hline\hline
    \end{tabular*}
    \end{table}

\begin{figure}[t]
    \centering
    \includegraphics[width=1.0\linewidth]{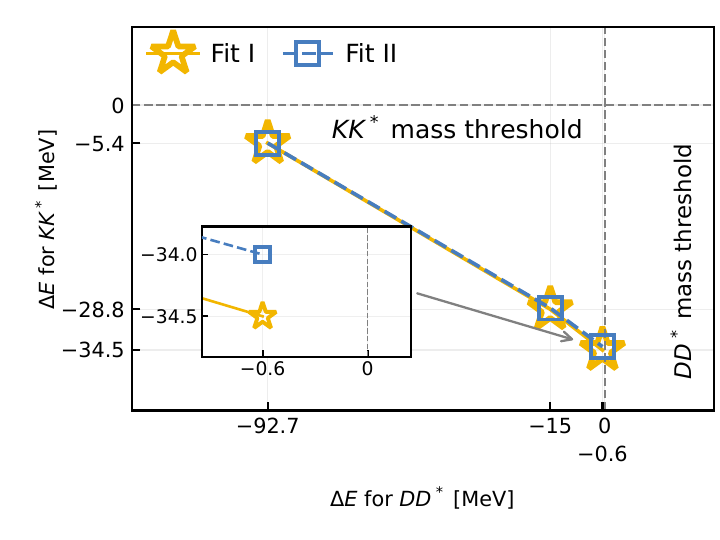}
    \caption{The $KK^*$ virtual-pole energy versus the $DD^*$ binding energy in the modified AL1 model for $g = 1.5, 1.0$, and $0.5$ (from left to right). }
    \label{fig:kkstar-pole-xi}
    \end{figure}

Beyond the numerical uncertainties in the few-body calculations, the inherent uncertainties of the underlying quark model itself can also be significant. For instance, the AL1 quark model predicts the $T_{cc}$ to be a bound state located approximately 15~MeV below the $DD^*$ threshold, whereas experimental observations~\cite{LHCb:2021auc,LHCb:2021vvq} identify it as a very shallow, near-threshold state. To investigate the impact of this model uncertainty, we modify the AL1 potential by scaling only the central Coulomb interaction between the two light quarks ($qq$, with $q=u,d$) by a factor $g$. Since the quark-antiquark interactions within the constituent mesons are unchanged, their masses and hence the meson-meson thresholds remain unchanged. We vary $g$ among 1.5, 1.0, and 0.5, and plot the pole positions of the $T_{ss}$ state against those of the $T_{cc}$ state in Fig.~\ref{fig:kkstar-pole-xi}. Comparing the results in Tables~\ref{tab:scattering-results} and \ref{tab:xi-half-results}, we find that reducing $g$ from 1.0 to 0.5 brings the $T_{cc}$ pole closer to the $DD^*$ threshold, while the virtual $KK^*$ pole moves farther from the $KK^*$ threshold. Specifically, for $g=0.5$, Table~\ref{tab:xi-half-results} gives a shallow $T_{cc}$ bound state located 0.6384~MeV below the $DD^*$ threshold, which is much closer to the experimental value, and a $T_{ss}$ virtual state located 34.49~MeV below the $KK^*$ threshold. 

We can see that, whether in the original AL1 model or the modified AL1 model that accounts for the shallower binding of the $T_{cc}$, the $T_{ss}$ state is consistently predicted to be a virtual state, with central pole positions approximately 30--35~MeV below the threshold. This conclusion differs significantly from those of chiral quark models and one-boson-exchange models, which typically predict  bound $T_{ss}$ states~\cite{Wang:2024kke,Ji:2024znj,Liu:2025fpx,Tang:2025kkstar}.

    \begin{table}[t]
    \caption{Low-energy properties of the $I(J^P)=0(1^+)$ channels in the modified AL1 quark model with $g=0.5$. The notation is the same as that in Table~\ref{tab:scattering-results}.
    }
    \label{tab:xi-half-results}
    \centering
    \setlength{\tabcolsep}{2pt}
    \begin{tabular*}{\columnwidth}{@{\extracolsep{\fill}}lcc@{}}
    \hline\hline
    Channel & $KK^*$ & $DD^*$  \\
    \hline
    $a_0$ (fm)
    & $-0.549^{+0.030}_{-0.030}$
    & $5.828^{+0.173}_{-0.165}$
   \\[3pt]
    $r_e$ (fm)
    & $2.358^{+0.379}_{-0.438}$
    & $0.9196^{+0.0027}_{-0.0023}$
     \\[3pt]
    BERW (MeV)
    & $-34.49^{+5.46}_{-8.15}$
    & $-0.6384^{+0.0401}_{-0.0424}$
    \\[3pt]
    Pole type & virtual & bound  \\
    Free space (MeV) & - & $-0.602$ \\
    \hline\hline
    \end{tabular*}
    \end{table}

      \section{Summary}
In this work, we have applied the artificial harmonic-oscillator (HO) trap method, previously used to study scattering in nuclear and atomic-molecular physics, to multiquark systems. This approach allows us to investigate hadron-hadron interactions and extract scattering phase shifts directly from the underlying quark-level interactions. To the best of our knowledge, this is the first application of this method to multiquark systems. It enables us to extend the capabilities of the quark model beyond the calculation of bound and resonance states, allowing for the systematic exploration of virtual states in multiquark systems as well.

As a concrete application, we investigated the doubly strange tetraquark with $I(J^P)=0(1^+)$ in the $KK^*$ channel. This state is the strange partner of the observed doubly charmed state and the widely anticipated doubly bottomed state. In our calculations, we carefully evaluated the systematic uncertainties and controlled them through a rigorous numerical setup. These include the basis truncation errors of the energy levels in the trap, the finite-$\omega$ corrections in the BERW formula, the truncation errors of the ERE, and the inherent theoretical uncertainties of the quark model itself. The reliability of our method is further validated by the fact that the binding energies extracted via the HO trap method for the $DD^*$ bound state and the first excited $BB^*$ bound state are in excellent agreement with those obtained from direct basis expansion calculations. Looking forward, this method can be broadly applied to the study of diverse hadronic interactions and the prediction of hadronic bound states, resonant states and virtual states.

Our calculations demonstrate that while there is an attractive interaction between the $K$ and $K^*$ mesons, it is not strong enough to form a bound state; instead, it generates a virtual state located 30--35~MeV below the $KK^*$ threshold. This conclusion is in stark contrast to many recent studies that predict bound $T_{ss}$ states~\cite{Wang:2024kke,Ji:2024znj,Liu:2025fpx,Tang:2025kkstar}. Although virtual states are generally more challenging to observe experimentally than bound states, their signatures can still be sought in specific processes. The $1^+$ doubly strange channel cannot strongly decay into two pseudoscalar kaons. In the present quark-model calculation, the $K^*$ is treated as stable and its decay $K^*\to K\pi$ is neglected; in nature, this decay provides a $KK\pi$ final state in which to search for the $T_{ss}$. The large $J/\psi$ sample collected at BESIII provides opportunities for such searches in $J/\psi$ decays~\cite{BESIII:2022jpsi}. Several experimental channels offer promising analysis settings, including $B^0\to K^-+[K^+K^{*0}]$~\cite{BABAR:2007kkstar}, $J/\psi\to K^-\bar K^0+[K^+K^{*0}]\to K^-\bar K^0K^+K^+\pi^-$~\cite{BES:2008phikkpi}, $pp\to K^+K^+\pi^-+X$~\cite{Tang:2025kkstar}, and the proposed $pn\to\Lambda\Lambda X$ reaction~\cite{Liu:2008ck}. The last reaction calls for quasi-free $pn\to\Lambda\Lambda X$ scattering on a deuteron or nuclear target at HIAF~\cite{An:2025lws}.

\section*{ACKNOWLEDGMENTS}
	We thank Xiao-Hai Liu for the helpful discussions. This project was supported by the National
    Natural Science Foundation of China (Grant No. 12605139 ), and  Start-up Funds of Southeast University (Grant No. 4007022506). This research work is supported by the Big Data Computing
Center of Southeast University.

    \section*{Data Availability Statement}

The data supporting the findings of this study are available in the Zenodo repository~\cite{zhang_2026_23395543} or can be obtained directly from the authors upon reasonable request.

\bibliography{ref.bib}

\end{document}

\title{Supplemental Material for\\
Doubly strange tetraquark as a $KK^*$ virtual state: quark-model study with harmonic traps}
	\author{Kai-Kai Zhang\,\orcidlink{0009-0003-3460-3483}}
    \affiliation{School of Physics, Southeast University, Nanjing 211189, China}
    \author{Zi-Yang Lin\,\orcidlink{0000-0001-7887-9391}}\email{linzy@seu.edu.cn}
    \affiliation{School of Physics, Southeast University, Nanjing 211189, China}
    \author{Yan-Ke Chen\,\orcidlink{0000-0002-9984-163X}}
	\affiliation{School of Physics and Center of High Energy Physics,
		Peking University, Beijing 100871, China}
	\author{Lu Meng\,\orcidlink{0000-0001-9791-7138}}\email{lmeng@seu.edu.cn}	
    \affiliation{School of Physics, Southeast University, Nanjing 211189, China}
\maketitle

\section{Meson masses and trapped spectra}\label{app:mesons}

\begin{figure*}[htbp]
	\centering
	\includegraphics[width=\textwidth]{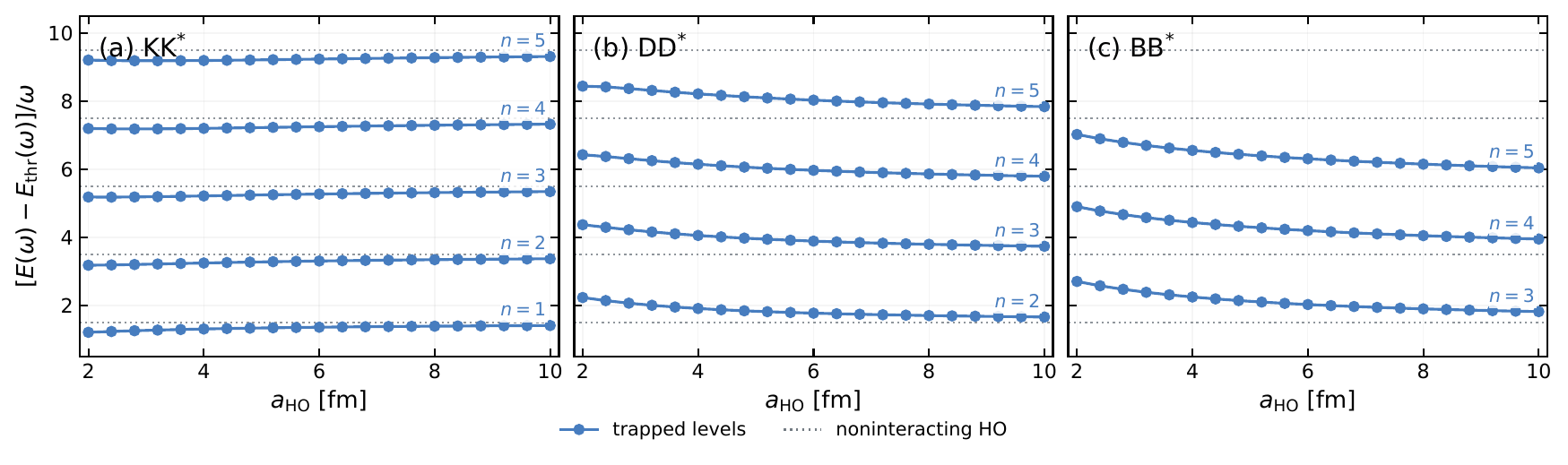}
	\caption{Trapped spectra of double strange, doubly charm and doubly bottom tetraquark states. Gray dotted lines denote the noninteracting relative-harmonic-oscillator values $2N+3/2$ for $N=0,\ldots,4$.}
	\label{fig:app-trap-spectra}
\end{figure*}

Table~\ref{tab:mesons} lists the AL1 meson masses, root-mean-square radii, and meson masses in the harmonic trap. Figure~\ref{fig:app-trap-spectra} shows the trapped spectra of the $KK^*$, $DD^*$, and $BB^*$ systems together with the noninteracting harmonic-oscillator levels.

\begin{table*}[htbp]
	\caption{Meson masses, root-mean-square radii and trapped internal energies in AL1 quark model. Experimental masses are isospin averaged~\cite{ParticleDataGroup:2026mpi}. Masses and energies are in MeV, and lengths are in fm.}
	\label{tab:mesons}
	\setlength{\tabcolsep}{4pt}
	\centering
	\begin{tabular}{llrrrrrrrr}
		\hline\hline
		& & & & & \multicolumn{5}{c}{Trapped energy at $a_{\rm HO}$ (fm)} \\
		$J^P$ & meson & $M_{\rm exp}$ & $M_{\rm AL1}$ & $\sqrt{\langle r^2\rangle}$
		& 2 & 4 & 6 & 8 & 10 \\
		\hline
		$0^-$ & $K$ & 495.6 & 490.9 & 0.5915
		& 491.38 & 490.97 & 490.95 & 490.95 & 490.94 \\
		& $D$ & 1867.3 & 1862.4 & 0.6113
		& 1862.51 & 1862.41 & 1862.41 & 1862.41 & 1862.41 \\
		& $B$ & 5279.6 & 5293.5 & 0.6258
		& 5293.51 & 5293.50 & 5293.50 & 5293.50 & 5293.50 \\
		$1^-$ & $K^*(892)$ & 893.7 & 903.6 & 0.8149
		& 904.38 & 903.61 & 903.56 & 903.56 & 903.55 \\
		& $D^*$ & 2008.6 & 2016.2 & 0.7018
		& 2016.32 & 2016.19 & 2016.18 & 2016.18 & 2016.18 \\
		& $B^*$ & 5324.8 & 5350.5 & 0.6604
		& 5350.56 & 5350.54 & 5350.54 & 5350.54 & 5350.54 \\
		\hline\hline
	\end{tabular}
\end{table*}

\section{Comparison of scattering parameters}
\label{sec:scattering-comparison}

Table~\ref{tab:scattering-comparison} compares our $DD^*$ and $BB^*$ scattering parameters with experimental and lattice results. The spread in the $DD^*$ scattering lengths reflects the sensitivity of $a_0$ to the position of a near-threshold pole.  In our sign convention, $1/a_0$ changes from negative to positive as an elastic $S$-wave virtual state becomes bound. When the pole reaches threshold, $1/a_0$ vanishes and the scattering length diverges.

For $BB^*$, both lattice results in the table are based on simulations at pion masses above the physical value. Aoki et al.~\cite{Aoki:2023nzp} extrapolated from $m_\pi\simeq410$, $570$, and $700$~MeV to $m_\pi=140$~MeV. Their result, $a_0=0.43(5)$~fm, is substantially smaller than the AL1 value of $2.72(4)$~fm. At $m_\pi=295$~MeV, Nagatsuka and Sasaki~\cite{Nagatsuka:2025szy} obtained $1/a_0=0.009(65)$~fm$^{-1}$ in our convention, consistent with the unitary limit.

In the AL1 model, both the $BB^*$ ground state and its first excitation are bound. The excited state lies only $1.7$~MeV below threshold and gives rise to the relatively large positive scattering length. Since AL1 overbinds the $T_{cc}$, it may also overestimate the attraction in the $T_{bb}$ channel. A weaker attraction could make the first excited state unbound.  More precise lattice determinations of the $BB^*$ scattering length near the physical pion mass would help constrain the quark-model interaction in this channel.

\begin{table*}[htbp]
\caption{
Comparison of the $S$-wave scattering length $a_0$ and effective
range $r_e$ for the $DD^*$ and $BB^*$ systems.
The quark-model and lattice results refer to the
$I(J^P)=0(1^+)$ channels, whereas the LHCb result is extracted
from the experimental line shape in the physical $D^0D^{*+}$ channel.
All results use the effective-range convention $p\cot\delta=-1/a_0+\frac{1}{2}r_e p^2+\cdots$.
Numbers in parentheses after the lattice references denote
$m_\pi$ in MeV. The modified AL1 results correspond to $g=0.5$.
}
\label{tab:scattering-comparison}
    \setlength{\tabcolsep}{0.5pt}
    \begin{tabular*}{\textwidth}{@{\extracolsep{\fill}}lcc@{}}
    \hline\hline
Source & $a_0$ (fm) & $r_e$ (fm) \\
\midrule
\multicolumn{3}{c}{$DD^*$} \\
\midrule
This work: AL1
  & $1.421^{+0.071}_{-0.070}$
  & $0.455^{+0.165}_{-0.146}$ \\
This work: modified AL1
  & $5.828^{+0.173}_{-0.165}$
  & $0.9196^{+0.0027}_{-0.0023}$ \\
LHCb~\cite{LHCb:2021auc}
  & \makecell[c]{
      $(7.16\pm0.51)$ \\
      $-i(1.85\pm0.28)$
    }
  & \makecell[c]{
      $-11.9<r_e\leq0$ \\
      (90\% C.L.)
    } \\
Padmanath et al.~\cite{Padmanath:2022cvl} (280)
  & $-1.04\pm0.29$
  & $0.96^{+0.18}_{-0.20}$ \\
Meng et al.~\cite{Meng:2023bmz} (280)
  & $-1.46\pm0.57$
  & $0.096\pm0.53$ \\
Lyu et al.~\cite{Lyu:2023xro} (146.4)
  & $\bigl(-0.050\pm0.054\bigr)^{-1}$\textsuperscript{$\dagger$}
  & $1.120^{+0.042}_{-0.085}$ \\
Nagatsuka et al.~\cite{Nagatsuka:2025szy} (295)
  & $-0.90\pm0.11$
  & $2.34\pm0.27$ \\
\midrule
\multicolumn{3}{c}{$BB^*$} \\
\midrule
This work: AL1
  & $2.721\pm0.043$
  & $1.065\pm0.023$ \\
Aoki et al.~\cite{Aoki:2023nzp} (140)
  & $0.43\pm0.05$
  & $0.18\pm0.06$ \\
Nagatsuka et al.~\cite{Nagatsuka:2025szy} (295)
  & $\bigl(+0.009\pm0.065\bigr)^{-1}$\textsuperscript{$\dagger$}
  & $1.808\pm0.082$ \\
\bottomrule
\bottomrule
\end{tabular*}
\begin{minipage}{\columnwidth}
\footnotesize
${}^{\dagger}$ Since the uncertainty intervals of $1/a_0$ include zero, we quote $a_0$ in reciprocal form.
\end{minipage}
\end{table*}

\bibliography{ref}